\documentclass[journal, 10pt]{IEEEtran}

\usepackage{amsthm,amsmath}
\usepackage{amsfonts}
\usepackage{amssymb}
\usepackage{mathrsfs}
\usepackage{mathtools}
\usepackage{float}
\usepackage[pdftex]{graphicx}
\usepackage{amsmath}
\usepackage{color}
\usepackage{multirow,multicol}
\usepackage{graphicx}
\usepackage{times}
\usepackage{textcomp}
\usepackage{verbatim}
\usepackage[table]{xcolor}
\usepackage{balance}
\usepackage{lipsum}
\usepackage[inline]{enumitem}
\usepackage{cuted}
\usepackage[caption=false,font=footnotesize]{subfig}
\usepackage{cite}
\usepackage{algpseudocode,algorithm}
\usepackage{hyperref}
\usepackage{tcolorbox}
\usepackage{setspace,epstopdf}
\usepackage[table]{xcolor}
\definecolor{color1}{RGB}{199,209,232}
\definecolor{color2}{RGB}{230,231,233}

\begin{document}
	
	\title{ A Survey of Deep Learning Architectures for Intelligent Reflecting Surfaces
	}
	\author{\IEEEauthorblockN{Ahmet M. Elbir, \textit{Senior Member, IEEE}, and Kumar Vijay Mishra, \textit{Senior Member, IEEE}}
		\thanks{A. M. Elbir is with the Department of Electrical and Electronics Engineering, Istinye University, 34396 Istanbul, Turkey and University of Luxembourg (e-mail: ahmetmelbir@ieee.org).} 
		\thanks{K. V. Mishra is with the United States DEVCOM Army Research Laboratory, Adelphi, USA (e-mail: kvm@ieee.org). }	
	}
	\maketitle
	
	\begin{abstract}
		Intelligent reflecting surfaces (IRSs) have recently received significant attention for  6G wireless communications as they enable the control of the wireless propagation environment. {The use of IRS also provides {reducing} the hardware complexity, physical size, weight as well as  cost of conventional large antenna arrays.} However, deployment of the IRS entails dealing with multiple channel links between the base station (BS) and the users. Further, the BS and IRS beamformers require a joint design, wherein the IRS elements must be rapidly reconfigured. Data-driven techniques, such as deep learning (DL), are critical in addressing these challenges. The lower computation time and model-free nature of DL make it robust against data imperfections and environmental changes. At the physical layer, DL has been shown to be effective for IRS signal detection, channel estimation, and active/passive beamforming using architectures such as supervised, unsupervised, and reinforcement learning. This article provides a synopsis of these techniques for designing DL-based IRS-assisted wireless systems.

	\end{abstract}
	%
	%
	

	\section{Introduction}
	\label{sec:Introduciton}
\begin{figure*}
		\centering
		{\includegraphics[width=0.8\textwidth]{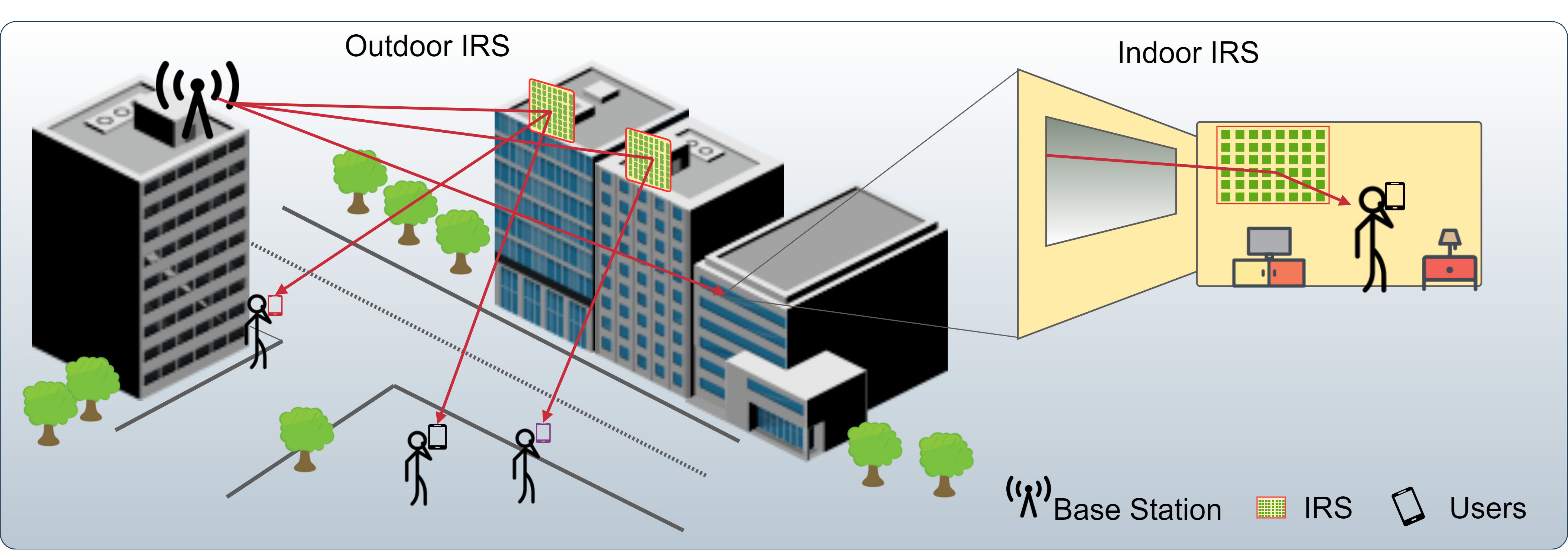} }
		\caption{IRS-assisted wireless communications for outdoor and indoor deployments. A BS on top of the infrastructure (left) communicates with the users on the ground through an intermediate IRS mounted on other buildings (center). The BS also serves users (right) inside the apartment building through an IRS placed on the wall of the room. 
		}
		\label{fig_IRS}
	\end{figure*}

	The sixth-generation (6G) millimeter wave (mm-Wave) massive multiple-input multiple-output (MIMO) systems require large antenna arrays with a dedicated radio-frequency (RF) chain for each antenna. This results in expensive and large system architectures that consume high power and processing resources. To reduce the number of RF chains while also maintaining sufficient beamforming gains, {hybrid analog and digital beamforming} architectures were introduced. However, the resulting cost and energy overheads using these systems remain a concern. {	Recently, intelligent reflective surfaces (IRSs) have emerged as a promising solution to enhance the spectrum and energy efficiency while constructing a favorable channel response by controlling the wireless propagation environment via large number of reconfigurable passive reflecting elements} (Fig.~\ref{fig_IRS}). {IRS is a feasible solution to implement low-cost and light-weight alternatives to large arrays complexity in both outdoor and indoor applications, usually with separate operating frequencies or spectral bands~\cite{irs_survey_2024_OJAP}. Thus, IRS is envisioned as one of the key enablers of smart electromagnetic environment (SEME) concept, wherein  the environment is no more an obstacle to wireless signals, but instead enables controlling and tailoring the propagation of electromagnetic (EM) waves.}

	

	An IRS is a two-dimensional (2D) reconfigurable metasurface that is composed of large periodic array of subwevelength scattering elements (meta-atoms). {The reflection coefficient of each IRS element is locally controlled via both phase and polarization such that the direction of the incoming EM wave is manipulated in real time, providing adaptive and programmable functionalities.} The phase shift is controlled via external signals by the base station (BS) through a backhaul control link. As a result, the incoming signal from the BS can be manipulated in real time, thereby, reflecting the received signal toward the users. Hence, the usage of IRS enhances the signal energy received by distant users and expands the coverage of the BS. It is, therefore, required to jointly design the beamformer parameters both at the IRS and BS. This achieves desired channel conditions, wherein the BS conveys the information to multiple users through the IRS~\cite{irs_survey_2024_OJAP,irs_secure_2025}.  Different from amplify-and-forward (AF) relay systems, an  IRS can have both active and passive components, which can provide a flexible configuration, thus, it has less active transmit modules or totally reflects the received signal as a passive surface. Thus, the IRS is much more energy- and  spectrum-efficient~\cite{irs_indoor_RIS_ISAC}.

	The accuracy of beamformer design strongly relies on the knowledge of the channel information. In fact, the IRS-assisted systems include multiple communications links, i.e., a direct channel from BS to users and a cascaded channel from BS to users through the IRS. This makes the IRS scenario even more challenging than the conventional massive MIMO systems. Furthermore, the wireless channel is dynamic and uncertain because of changing IRS configurations. Consequently, there exists an inherent uncertainty stemming from the IRS configuration and the channel dynamics. These characteristics of IRS make the system design very challenging~\cite{irs_RL_energyEfficient_,irs_RL_BF_IRSonly}.
	
	To address the aforementioned uncertainties and non-linearities imposed by channel equalization, hardware impairments, and sub-optimality of high-dimensional problems, model-free techniques have become common in wireless communications~\cite{irs_survey_2024_OJAP}. \textcolor{black}{In this context, deep learning (DL) is particularly powerful in extracting the features from the raw data and providing a ``meaning'' to the input by constructing a model-free data mapping with a huge number of learnable parameters.} Furthermore, DL is helpful when modeling the channel characteristics thanks to its data-driven structure. As listed below, DL is more efficient than model-based techniques that largely rely on mathematical models:
	\begin{itemize}
		\item A learning model constructs a non-linear mapping between the raw input data and the desired output to approximate a problem from a model-free perspective. Thus, its prediction performance is robust against the corruptions/imperfections in the wireless channel data.
		\item DL learns the feature patterns, which are easily updated for the new data and adapted to environmental changes. \textcolor{black}{In the long run, this results in lower computational complexity than a model-based optimization~\cite{irs_RL_BF_IRSonly}}.
		\item DL-based solutions have significantly reduced run times because of parallel processing capabilities. On the other hand, it is not straightforward to achieve parallel implementations of conventional optimization and signal processing algorithms~\cite{elbir2020_FL_CE}. 
	\end{itemize}
	The aforementioned advantages have led to DL superseding the optimization-based techniques in the system design for the physical layer of the wireless communications.

	Lately, the IRS-aided wireless systems have exploited DL to handle very challenging problems. \textcolor{black}{For instance, signal detection in IRS requires development of end-to-end learning systems under the effect of channel and beamformers~\cite{irs_DL_detection}. The channel needs to be estimated for multiple communication links, i.e., BS-user and BS-IRS-user~\cite{elbir_LIS}. Finally, beamformers are designed (by solving complex optimization problems) for phase shifters at both BS and passive elements of the IRS~\cite{lis_channelEst_reflectedBFDesign}. The DL-based techniques are able to handle the multidimensional, huge datasets in all these problems and may also be employed for channel modeling~\cite{irs_survey_2024_OJAP}, where the conventional model-based approaches are not very useful.} There have been recent surveys on applying DL~\cite{dl_6G_survey} and IRS~\cite{irs_survey_2024_OJAP} individually to wireless communications. In this article, we provide an overview of systems that jointly employ both approaches. In particular, we describe DL techniques (Table~\ref{tableSummary}) for three main IRS problems: signal detection, channel estimation, and beamforming. Each of these requires different DL architectures, which have so far included supervised learning (SL), unsupervised learning (UL), reinforcement learning (RL) and federated learning (FL). \textcolor{black}{The UL and RL do not require labeling; SL needs a labeled dataset; and} {FL has a distributed structure for model training, which can be performed via either SL, UL or RL.} We provide a detailed synopsis of the advantages and shortcomings of each algorithm for these three applications in the subsequent sections. \textcolor{black}{We also discuss the design challenges in terms of DL perspective and the emerging IRS-assisted systems for THz communications, cell-free networks and integrated sensing and communications (ISAC), and highlight the related future research directions.}

	


	%
	%
	%
	%
	%


	{

	}

	\begin{table*}
		\caption{DL-based Techniques for IRS-assisted Wireless Systems
		}
		\label{tableSummary}
		\centering
		\begin{tabular}{p{0.06\textwidth}p{0.2\textwidth}p{0.33\textwidth}p{0.32\textwidth}}
			\hline 
			\hline
			\bf  Scheme \cellcolor{color1}
			&\cellcolor{color2}\bf NN Architecture
			& \bf Benefits\cellcolor{color1}
			&\cellcolor{color2}\bf Drawbacks 
			\\
			\hline 
			\multicolumn{4}{c}{\bf Signal detection}\\
			\hline
			SL~\cite{irs_DL_detection}\cellcolor{color1}
			&\cellcolor{color2}	 MLP with $3$ layers 
			& \cellcolor{color1}  No need for channel estimation algorithm
			&Still needs to design beamformers and requires huge datasets and deeper NN architectures \cellcolor{color2} 
			\\
			\hline
			\multicolumn{4}{c}{\bf  Channel estimation}\\
			\hline
			SL~\cite{elbir_LIS}\cellcolor{color1}
			&	\cellcolor{color2} Twin CNNs with  $3$ convolutional, $3$ fully connected layers
			&\cellcolor{color1} Each user estimates its own channel with the trained model
			&\cellcolor{color2} Data collection requires channel training by turning on/off each IRS element
			\\
			\hline
			\cellcolor{color1}
			FL \cite{elbir2020_FL_CE}
			&\cellcolor{color2}	 A single CNN with $3$ convolutional, $3$ fully connected layers
			&\cellcolor{color1} Less transmission overhead for training, A single CNN estimates both cascaded and direct channels
			& Performance depends on the number of users and the diversity of the local datasets\cellcolor{color2}
			\\
			\hline
			SL\cite{irs_deepDenoisingNN_CE}	\cellcolor{color1}
			&\cellcolor{color2}	DDNN with $15$ convolutional layers
			&\cellcolor{color1} Leverages both compressed sensing (CS)
			and DL methods
			&  \textcolor{black}{Requires active IRS elements. High prediction complexity arising from CS algorithms}\cellcolor{color2}
			\\
			\hline
			\multicolumn{4}{c}{\bf Beamforming}\\
			\hline
			\cellcolor{color1}
			SL \cite{lis_channelEst_reflectedBFDesign}
			&\cellcolor{color2}	MLP with $4$ layers 
			&\cellcolor{color1}   Reduced pilot training overhead
			&\cellcolor{color2} \textcolor{black}{Requires active IRS elements} for channel training
			\\
			\hline
			\cellcolor{color1}
			RL \cite{irs_BF_RL_standalone_alkhateeeb} 
			&\cellcolor{color2}	 DQN with $4$ layers
			&\cellcolor{color1}  Provides standalone operation since RL does not require labels like SL
			&\cellcolor{color2}  Longer training. \textcolor{black}{Active IRS elements needed for channel acquisition}
			\\
			\hline
			\cellcolor{color1}
		RL \cite{irs_RL_BF_IRSonly}
		&\cellcolor{color2} DDPG	with $4$-layered actor and critic networks
		&\cellcolor{color1}  Better performance than DQN  
		&\cellcolor{color2}Large number of NN parameters are involved 
		\\
		\hline
		FL \cite{irs_FL_BF_fromChannel}\cellcolor{color1}
		&	\cellcolor{color2} MLP with $6$ layers
		&\cellcolor{color1}  Less transmission overhead 
		during model training
		&\cellcolor{color2} IRS must be connected to the PS
		\\
		\hline
		\multicolumn{4}{c}{\bf Secure beamforming}\\
		\hline
		RL \cite{irs_secure_2025}\cellcolor{color1}
		&\cellcolor{color2}	DDPG
		&\cellcolor{color1}Robust against eavesdropping
		&\cellcolor{color2} High model training complexity
		\\
		\hline
		\multicolumn{4}{c}{\bf Energy-efficient beamforming}\\
		\hline
		RL \cite{irs_RL_energyEfficient_}\cellcolor{color1}
		&\cellcolor{color2}	 DQN
		&\cellcolor{color1}Energy-efficient and robust against uncertainties
		&\cellcolor{color2} IRS beamforming only  
		\\
		\hline
		\multicolumn{4}{c}{\bf Indoor beamforming}\\
		\hline
		SL \cite{irs_indoor_RIS_ISAC}\cellcolor{color1}
		&\cellcolor{color2}	 CNN with $5$ layers
		&\cellcolor{color1}Reduces hardware complexity of multiple BSs and improves RSS 
		for indoor environments
		&\cellcolor{color2} Learning model performance relies on room conditions
		\\
		\hline
		\hline 
	\end{tabular}
\end{table*}


\begin{figure*}
	\centering
	{\includegraphics[width=.8\textwidth]{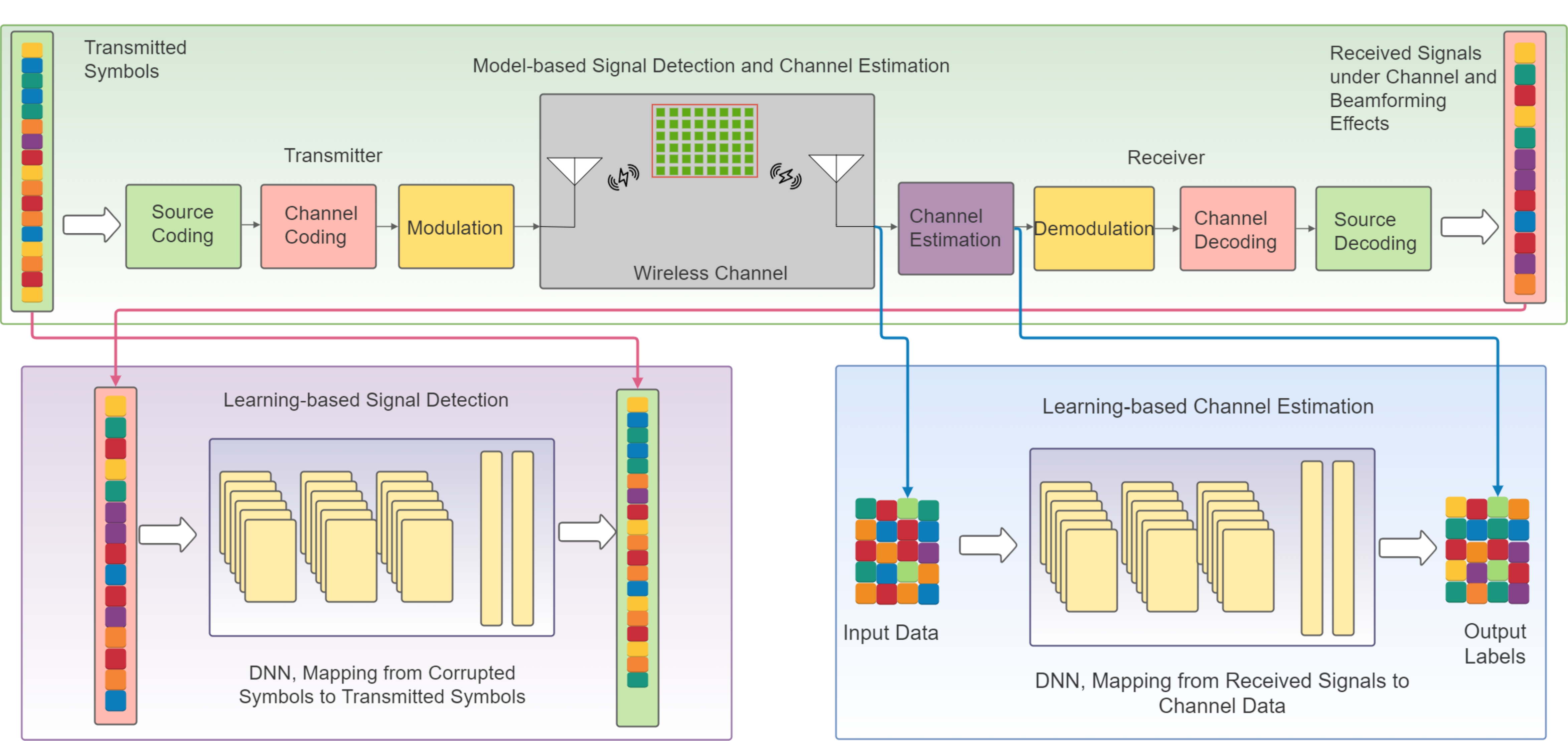}}
	\caption{Model-based versus learning-based frameworks for signal detection and channel estimation. Model-based approach (top) comprises multiple subsystems to process the received signal. Learning-based signal detection (bottom, left) provides an end-to-end data mapping from the corrupted symbols under the channel effects at the receiver to the transmit symbols. Learning-based channel estimation (bottom, right) maps the input received signals to the channel estimate as output labels. 
	}	
	\label{fig_SD_CE}
\end{figure*}

	\section{DL-Based Signal Detection in IRS}
	
	
	To leverage DL for signal detection, \cite{irs_DL_detection} devised a multi-layer perceptron (MLP) for mapping the channel and reflecting beamformer data symbols to the transmit symbols. The MLP is a feedforward neural network (NN) composed of multiple hidden layers. The framework in \cite{irs_DL_detection} uses three fully connected layers. {\color{black}Once the MLP is trained on a dataset composed of received-transmitted data symbols for Rayleigh fading channels, each user feeds the learning model with the block of received symbols.} These blocks account for the effect of channel and beamformers. Then, MLP yields the estimated transmit symbols.
	
	A major advantage of this approach is its simplicity in that the learning model estimates the data symbols directly, without a prior stage for channel estimation. Thus, this method is helpful in reducing the cost of channel acquisition.
	In~\cite{irs_DL_detection}, a bit-error-rate (BER) analysis has shown that the DL-based IRS signal detection (DeepIRS) provides better BER than the minimum mean-squared-error (MMSE) and close performance to the maximum likelihood estimator.
	However, a few challenges remain to achieve a reliable performance. The training data should be collected under several channel conditions and different beamformer configurations so that the trained model learns the environment well and reflects accurate performance in different scenarios. This is a particularly challenging task because it requires collection of the training data for different user locations. In conclusion, DL-based signal detection is helpful for bypassing the channel estimation stage. However, this may require huge training datasets collected under different channel conditions. \textcolor{black}{An alternative is to consider estimating the wireless channel via DL, as discussed in the next section.  }


	\section{DL-Based IRS Channel Estimation}
	{During channel acquisition, the BS transmits pilot signals in the downlink, which are received and processed at the user. Using the pilot signals, which are known as a priori, the user estimates the impact of the wireless channel on these pilots and transmits the channel state information (CSI) back to BS via uplink. In IRS-assisted scenario, channel estimation is even more challenging as the user needs to estimate CSI over multiple links, e.g., BS-IRS, IRS-user and BS-user. The complexity also increases with the number of elements in the IRS.} A common approach is to turn on and off each individual IRS element one by one while also using orthogonal pilot signals to estimate the channel between the BS and the users through the IRS.  The SL approach proposed in \cite{elbir_LIS} estimates both direct and cascaded channels via twin convolutional neural networks (CNNs). First, the received pilot signals at the user are collected by sequentially turning on the individual IRS elements. Then, the collected data are used to find the LS estimate of the cascaded and the direct channels. Both CNNs are trained to map the LS channel estimates to the true channel data. The upshot is that each user estimates its own channels only once and feeds the received pilot data (LS estimate) to the trained CNN models. The CNNs have higher tolerance than MLP against the channel data uncertainties, and imperfections (such as switching mismatch) of IRS elements.
	

	\textcolor{black}{When the model training is conducted at the user with huge datasets as in~\cite{elbir_LIS}  for various channel/user/configurations, the system may lack sufficient computational capability.} This is overcome by FL-based training \cite{elbir2020_FL_CE}, where the learning model updates are computed at the devices (nodes) and aggregated at the BS (central server), thereby eliminating the transmission of raw data.   FL significantly reduces the transmission overhead since the size of the datasets is usually larger than the size of the learning model, {\color{black}and its performance improves as the number of users increases~\cite{elbir2020_FL_CE,irs_FL_BF_fromChannel}}. {\color{black}Furthermore, instead of using two CNNs demanding two datasets as in~\cite{elbir_LIS}, a single CNN in~\cite{elbir2020_FL_CE} jointly estimates both cascaded and direct channels.}
	
	\begin{figure}[t]
		\centering
		{\includegraphics[draft=false,width=\columnwidth]{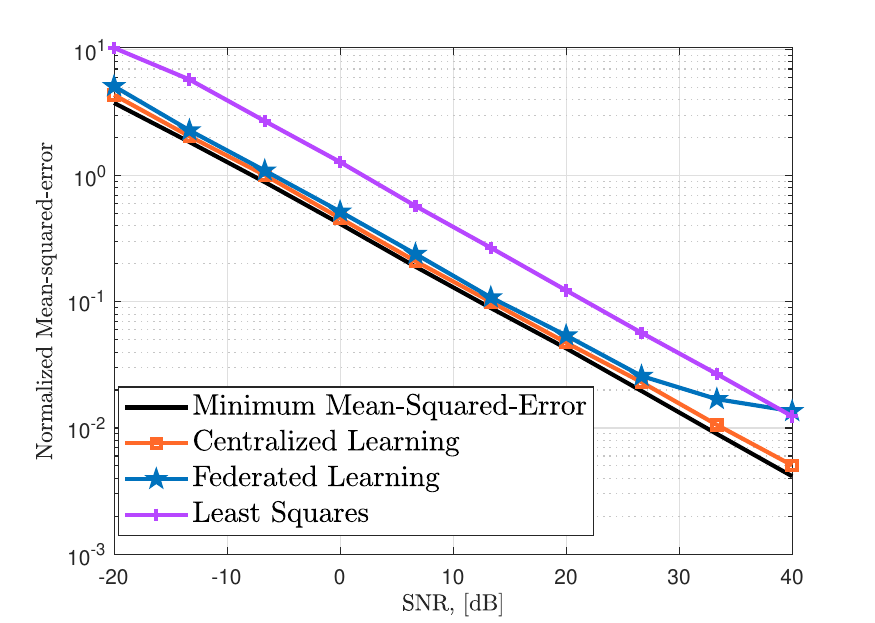} } 
		\caption{The mean-squared error of channel estimates normalized against ground truth channel, obtained using CNN in centralized and federated learning frameworks, MMSE and LS. The BS consists of $64$ antennas and IRS employed $64$ passive reflecting elements~\cite{elbir_LIS,elbir2020_FL_CE}.
		}
		\label{fig_NMSE_ALL}
	\end{figure}
	
	Although FL reduces the transmission overhead during model training, its training performance is upper bound by the centralized model training, i.e., training the model with the whole dataset at once. Therefore, the prediction performance of FL is usually poorer than that of centralized learning (CL). As shown in Fig.~\ref{fig_NMSE_ALL}), CL and FL frameworks are compared with the MMSE and the LS estimation in terms of normalized mean-squared error (NMSE). We note that FL performs slightly poorer than CL in high SNR regimes. Despite this, FL significantly reduces the transmission overhead, e.g., approximately ten-fold reduction in the number transmitted symbols~\cite{elbir2020_FL_CE}. 
	

	{SL-based channel estimation techniques suffer from high channel training overhead.} In this context, compressive channel estimation with deep denoising neural networks (DDNNs) is very effective \cite{irs_deepDenoisingNN_CE}. It employs a hybrid passive/active IRS architecture, where the active IRS elements are used for uplink pilot training and passive ones for reflecting the signal from the BS to the users. Once the BS collects the compressed received pilot measurements, the complete channel matrix is recovered through sparse reconstruction algorithms such as orthogonal matching pursuit (OMP). Then, DDNN is used to improve the channel estimation accuracy by exploiting the correlation between the real and imaginary parts of the mm-Wave channel in the angular-delay domain. During training, the input is the OMP-reconstructed channel matrix and the output is the noise, i.e., the difference between the OMP estimate and the ground truth channel data.	This method leverages both CS and DL yielding a performance better than using these techniques individually. The major drawback is the additional hardware complexity introduced by the active IRS elements.
	Furthermore, the OMP algorithm is used in place of the raw received pilot measurements for constructing the input. This requires repeated execution of the OMP algorithm thereby increasing the prediction complexity over the DL methods in~\cite{elbir_LIS} and~\cite{elbir2020_FL_CE}. The additional OMP stage in DDNN aids in achieving lower MSE than the DDNN-only architectures. 
	
	
	%
	
	\section{DL-Aided Beamforming for IRS Applications}
	The aim of the beamforming design is to  maximize the spectral efficiency (SE) of the IRS-aided system. 
In the following, we present various beamforming applications in IRS-assisted systems. 


\subsection{Beamforming at the IRS}
The IRS beamforming requires passive elements continuously to reliably reflect the BS signal to the users.	Here, the MLP architecture \cite{lis_channelEst_reflectedBFDesign} is helpful in designing the reflect beamforming weights using active IRS elements \cite{irs_deepDenoisingNN_CE}. These elements are randomly distributed through the IRS. They are used for pilot training, after which compressed channel estimation is carried out using OMP. \textcolor{black}{In order to collect the dataset, the reflect beamforming weights are optimized by using the estimated channel data. Finally, a training dataset is constructed with channel data and reflect beamformers as the input-output pairs for an SL framework.} Note that the active IRS elements present similar shortcomings as in~\cite{irs_deepDenoisingNN_CE}. However, the method in~\cite{lis_channelEst_reflectedBFDesign} excels by leveraging DL for designing beamformers. Thus, an interesting future work can be the combination of both DL-based approaches~\cite{irs_deepDenoisingNN_CE} and~\cite{lis_channelEst_reflectedBFDesign} for joint channel estimation and beamforming via DL.


The labeling process in~\cite{lis_channelEst_reflectedBFDesign} demands solving an optimization problem for each channel instance in the training data generation stage. One possible way to mitigate this is to use label-free techniques, such as UL.	The UL approach in \cite{irs_indoor_RIS_ISAC} for reflect beamforming design employs a CNN with five convolutional layers. {The loss function to be minimized is selected as the negative of the channel gain so that the similarity between the channel instances are measured.} However, this technique yields the phase information at the output uniquely for each training sample. Consequently, the beamformers implicitly behave like a label in the training process. 



\begin{figure}[t]
	\centering
	{\includegraphics[width=\columnwidth]{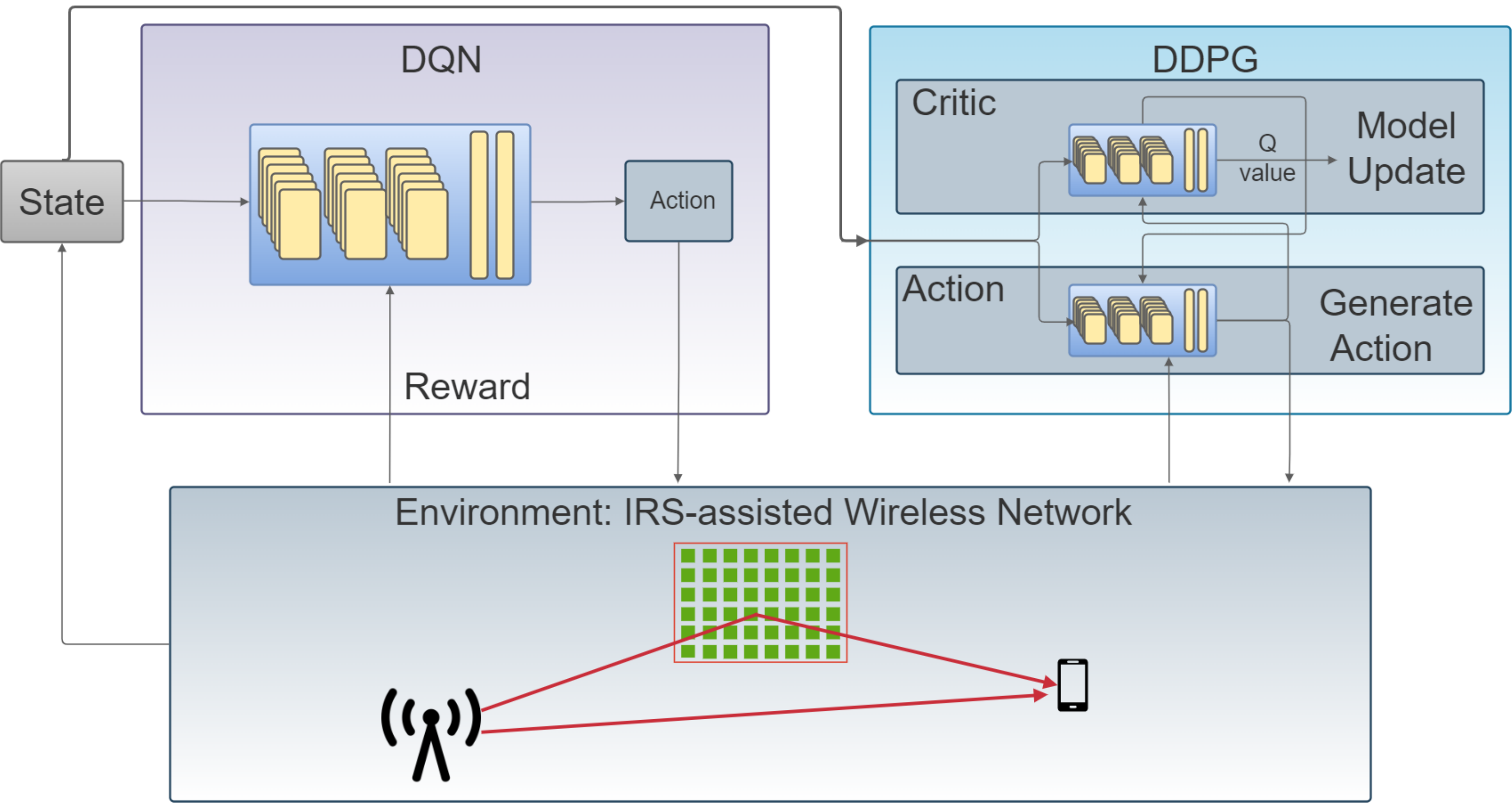} }
	\caption{In RL, the DQN and DDPG architectures accept the same state (channel data and received SNR) and environment data (beamformers to be evaluated). The DQN involves training a single neural network based on the reward determined by the environment. On the other hand, the DDPG has multiple neural networks, where actor-critic architectures are used to compute actions and target values, respectively. 
	}
	\label{fig_DQN_DDPG}
\end{figure}

In order to eliminate the expensive labeling process of the SL-based techniques, \cite{irs_BF_RL_standalone_alkhateeeb} employed RL to design the reflect beamformers for single-antenna users and BS. The RL is a promising approach that directly yields the output by optimizing the objective function of the learning model. First, the channel state is estimated by using two orthogonal pilot signals. An action vector is selected either by exploitation (using the prior experience of the learning model) or exploration (using a predefined codebook). After computing the achievable rate based on the selected action vector from the environment, a reward or penalty is imposed by comparing with the achievable rate with a threshold. Upon reward calculation, a Deep Quality Network (DQN) (Fig.~\ref{fig_DQN_DDPG}) updates the map from the input state (channel data) to the output action (action vector composed of reflect beamformer weights). \textcolor{black}{The training data is generated in an EM simulation tool}, and this process is repeated for several input states until the learning model converges. \textcolor{black}{While RL is not an IRS-specific technique, it is particularly useful in lowering the overhead of the labeling process as compared to \color{black} SL architectures deployed by RNN or CNN models, which require labeled datasets.}  The RL algorithm learns to reflect beamformer weights based on the optimization of the achievable rate. Thus, RL presents a solution for online learning schemes, where the model effectively adapts to the changes in the propagation environment. However, RL techniques have longer training times than the SL approaches because the reward mechanism and discrete action spaces make it difficult to reach the global optimum. The label-free process implies that the RL usually has a slightly poorer performance than the SL.

%
%

To accelerate the training stage by the use of continuous action spaces, a deep deterministic policy gradient (DDPG) (Fig.~\ref{fig_DQN_DDPG}) was introduced in \cite{irs_RL_BF_IRSonly}. Here, actor-critic network architectures are used to compute actions and target values, respectively. 	First, the learning stage is initialized by the use of an input state excited by cascaded and direct channels. Given the state information, a deep policy network (DPN) (actor) constructs the actions (reflection beamformer phases). \textcolor{black}{Here, the DPN provides a continuous action space that converges faster than the DQN architecture in~\cite{irs_BF_RL_standalone_alkhateeeb}.} The action vector is used by the critic network architecture to estimate the received signal-to-noise ratio (SNR) as an objective. This SNR then yields the target beamformer vector under the learning policy. Using the gradient of DPN, the network parameters are updated and the next state is constructed as the combination of the received SNR and the reflecting beamformers. 	This process is repeated until it converges. 

An additional benefit of this approach is that it outperforms fixed-point iteration (FPI) algorithms used to solve reflect beamforming optimization. Moreover, the continuous action space representation with DPN in DDPG provides robustness of the learning model against changes in channel data. 
However, multiple NN architectures (actor and critic networks) increase the number of learning parameters and aggravate model update requirements for each architecture.


Even if RL is a label-free approach that reduces the overhead during training data generation, training approaches in~\cite{irs_RL_BF_IRSonly,irs_BF_RL_standalone_alkhateeeb} demand expensive transmission overhead to be trained on huge datasets. This is mitigated in FL techniques.	The FL approach in~\cite{irs_FL_BF_fromChannel} learns the IRS reflect beamformers by training an MLP by computing the model updates at each user with the local dataset. The model updates are aggregated in a parameter server (PS), which is connected to the IRS. The MLP input is the cascaded channel information and the output labels are IRS beamformer weights. The federated architecture lowers the transmission overhead during training. However, it is assumed that the PS is connected to the IRS. The simple architecture of the IRS could make this infeasible. It is more practical to access the PS via BS for model training.

\subsection{Secure Beamforming}

The RL-based secure beamforming \cite{irs_secure_2025} minimizes the secrecy rate by jointly designing the beamformers at the IRS and BS to serve multiple legitimate users in the presence of eavesdroppers. The RL algorithm accepts the states as the channel information of all users, secrecy rate, and transmission rate. Similar to~\cite{irs_RL_BF_IRSonly}, the action vectors are beamformers at the BS and IRS. The reward function is designed based on the secrecy rate of users. 
A DDPG is trained to learn the beamformers by minimizing the secrecy rate while guaranteeing the quality-of-service requirements. The model training takes place at the BS, which is responsible for collecting the environment information (channel data) and making decisions for secure beamforming. This scheme is more realistic and reliable than that of~\cite{irs_BF_RL_standalone_alkhateeeb,irs_RL_BF_IRSonly}, which ignores the effect of eavesdroppers. The learning model includes high-dimensional state and action information, such as the channels of all users and beamformers of BS and IRS. This may necessitate more computing resources for training than non-secure IRS~\cite{irs_BF_RL_standalone_alkhateeeb,irs_RL_BF_IRSonly} and conventional SL techniques~\cite{elbir_LIS,lis_channelEst_reflectedBFDesign}.

\subsection{Energy-Efficient Beamforming}
The IRS configuration dynamically changes depending on the network status. It is very demanding for the BS to optimize the transmit power every time the on/off status of IRS elements is updated. This could be addressed by accounting for energy efficiency in the beamformer design problem. In~\cite{irs_RL_energyEfficient_}, a self-powered IRS scenario maximizes energy efficiency by optimizing the transmit power and the IRS beamformer phases. In this DQN-based RL approach, the BS learns the outcome of the system performance while updating the model parameters. Thus, the BS makes decisions to allocate the radio resources by relying on only the estimated channel information. \textcolor{black}{The dataset for the RL framework has states selected as the estimated channels from users and the energy level of the IRS. Meanwhile, the action vector includes the transmit power, the IRS beamformer phases, and the on/off status of the IRS elements.} The learning policy is based on the reward which is selected as the energy efficiency of the overall system. In contrast to other beamforming schemes~\cite{elbir_LIS,lis_channelEst_reflectedBFDesign,irs_RL_BF_IRSonly}, the major advantage of this approach is taking into account the energy efficiency of the overall system. 
However, this work considers only IRS beamforming and ignores the same at the BS.

\subsection{Beamforming for Indoor IRS}
Different from the above scenarios, \cite{irs_indoor_RIS_ISAC} addresses the IRS beamformer design problem in an indoor communications scenario to increase the received signal strength (RSS) (see Fig.~\ref{fig_IRS}). This is particularly useful for indoor sensing applications as well as from  the perspective of low hardware complexity because it employs lightweight CNN models is employed. The CNN architecture in \cite{irs_indoor_RIS_ISAC} accepts the IRS channel data as input and it yields the IRS beamformer phases at the output in unsupervised manner. 	The learning model trains on specific room environments and may perform poorly for different room conditions or different obstacle distributions in the same room. This is mitigated in RL-based solutions which are highly adaptive to different environments~\cite{irs_BF_RL_standalone_alkhateeeb,irs_RL_BF_IRSonly}.
As a result, among the various DL-based beamforming methods, RL does not involve labeling and, in this sense, it is more advantageous than SL. Further, DQN is an RL technique that has a simpler architecture with longer convergence times than, say, DDPG which consists of actor-critic spaces and provides better accuracy and faster convergence. Since FL offers reduced transmission overheads, an integrated FL-RL framework may be considered in the future.


\section{Challenges and Future Research Directions}
Several challenges remain for DL architectures to reach their full potential in realizing significant performance gains and efficiency for IRS-assisted wireless systems. 	In the following, we separately discuss the design challenges of DL models and IRS-assisted emerging applications.


%


\subsection{DL-Related Challenges}
The need for massive data collection and subsequent training is a bottleneck in the successful implementation of DL-based techniques for all wireless communications tasks: signal detection, channel estimation, and beamformer design.

\subsubsection{Data collection} 
{Massive data collection hampers the successful performance of DL-based techniques for all wireless communications tasks: signal detection, channel estimation, and beamformer design. The collected dataset should represent the different propagation environments for accurate performance. To overcome this challenge, the data collection should be performed for various scenarios with different user locations/directions, channel gains, number of paths, etc. Another challenging issue in data collection is labeling. This requires solving an optimization problem for each of the data samples depending on the application. For instance, the active/passive beamformer weights are the labels and they should be each data samples. e.g., CSI data. Labeling method has a direct impact on the accuracy of the learning model. Because, the learning accuracy is upperbounded by the performance of the labeling method. Thus, efficient but accurate labeling methods should be employed.} 


\subsubsection{Model training}
The models are usually trained offline prior to their online deployment at a PS connected to the BS. \textcolor{black}{In addition, the model training complexity increases with the number of IRS elements and IRSs deployed between the users and the BS.}  \textcolor{black}{Some experimental studies include IRS with $10000$ elements operating at $10.5$ GHz~\cite{irs_survey_2024_OJAP}. } 	\textcolor{black}{Hence, a huge transmission overhead is introduced for model training. 	The FL has the potential to reduce this cost and enable communication-efficient model training (see, e.g., Fig.~\ref{fig_NMSE_ALL}).} 
{Here, the integration of the DL architectures may provide leveraging their the complementary strength. For instance, FL is useful for reducing the transmission overhead during training while RL provides label-free model training. Thus, the combination of FL- and RL-based learning policies not only exhibits a communications-efficient model training but also provides environmental adaptation in IRS applications. }	{Furthermore, the hybridization of the DL schemes can also present a trade-off between computational complexity and learning accuracy\cite{hfcl_TSP}. For instance, a hybrid centralized and federated strategy could be useful as FL requires computational power at the edge users while the model is trained at the PS in the centralized scheme. Thus, a portion of the users who do not have the computational resources for model training can send their dataset to the PS while the remaining users perform training on their own via FL.   }

\subsubsection{Environment adaptation and robustness} 
{The behavior of the channel affects all DL-based tasks including channel estimation, beamforming, user scheduling, power allocation, and antenna selection/switching.
	Addressing the trade-off between the bias and the variance of the model output is essential for robust performance. This is usually achieved through data validation so that the learning model does not either over-fit or under-fit the training data. Nonetheless, this does not generalize the learning performance to different environments. Thus, the learning model should be both robust to the imperfections in the data and generalizable for various scenarios. While collecting very large amount of data is useful, it may reduce the learning accuracy due to having too many similar data samples for different cases. To this end, online learning techniques can be devised, wherein the learning model adapts to the new incoming data by updating its model parameters. Furthermore,  wider and deeper learning models are required to cover larger data spaces and provide a robust performance against the changes in the environment.}


{\color{black}
	\subsection{IRS-Related Challenges}
	\textcolor{black}{ Specific implementation challenges have also been identified within emerging technologies, some of which we elaborate on here.}\\ 

	{
		\subsubsection{Hardware Constraints} The implementation of DL-based IRS involves several hardware challenges. For instance, the complexity of the channel training scales with the number of IRS elements. This increases the complexity of both the channel acquisition and model training. This is especially challenging in large-scale IRS deployments as most of the IRS channel acquisition methods involve turning on/off the each IRS elements one by one during channel training. Likewise, it causes processing of huge data for inference. One way to overcome this issue is to take into account the redundancy in the IRS training data to reduce the complexity~\cite{elbir_LIS}. Another approach may be employing partitioning the IRS elements during training as subarrays. While this approach can provide significant reduction in the complexity, it leads to performance degradation.
		
	}
	
	Another hardware constraint is that the IRS elements introduce discrete phase shifts as it is costly to manufacture each reflecting element with infinite-level resolution~\cite{irs_DRL_Abdallah2024Jan}. The error due to discrete phase shifts impacts the reflection coefficient of the IRS, i.e.,  the ratio between the refracted and the incident electric field. To account for these imperfections, DL-based solutions can be helpful. Fig.~\ref{fig_RL_BF} shows the beamforming gain performance for multi-agent RL algorithm, wherein the beampattern for BS and IRS are jointly designed by employing an IRS reflection codebook based with discrete phase shifts. We see that RL-based beamforming provides higher gain than the use of DFT codebooks.

	{
		\subsubsection{Power Consumption and Latency}
		The IRS-aided systems have shown to provide higher SE with less power consumption thanks to extending the wireless coverage and employing less active elements~\cite{irs_survey_2024_OJAP}. Nevertheless, it involves training complexity due to large number of IRS elements, for which DL-based solutions can be helpful to reduce the real-time computational latency~\cite{irs_deepDenoisingNN_CE,irs_BF_RL_standalone_alkhateeeb}. Compared to optimization-based approaches, the advantage of DL-based techniques is to employ neural networks with parallel processing capability. This is especially important to achieve low latency requirements for 6G, e.g., less than 1 ms. It has been shown that the DL-based techniques have achieved successful resource allocation for 6G applications within every transmission interval of 0.125 to 1 ms~\cite{dl_6G_survey}. In the meantime, advanced hardware capabilities are highly important to enable DL in mobile communications to support training and inference processes within the real-time computational constraints. Furthermore, the real-time implementation may also require the re-training the learning model or update a portion of its parameters. This may cause additional overhead from computational perspective, and it should be also taken into account for real-time implementation.	
		
	}
	
	
	\begin{figure}[t]
		\centering
		{\includegraphics[width=.5\textwidth]{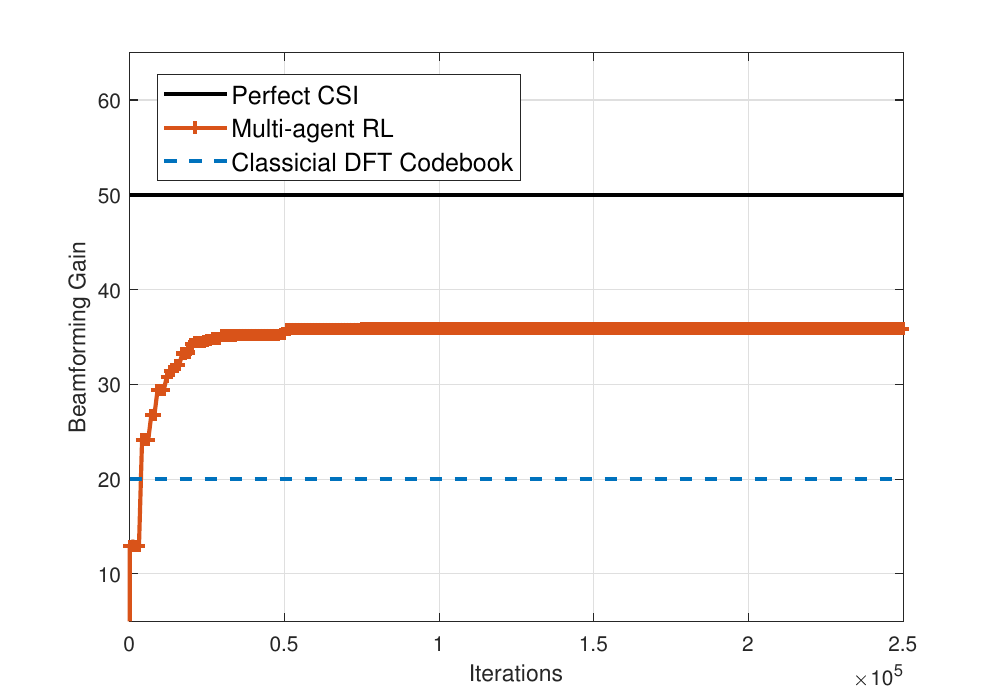} }
		\caption{Beamforming gain for IRS with discrete phase shifts, when $L=32$.
		}
		\label{fig_RL_BF}
	\end{figure}

	\subsubsection{THz implementation} 
	Compared to mm-Wave, the propagation loss is more significant in THz bands thereby leading to shorter ranges. While the mm-Wave channel model is based on a single line-of-sight (LoS) path with several non-LoS paths, the THz channel is largely a superposition of multiple LoS paths. Here, DL-aided IRS helps in extending the BS coverage. Apart from the challenges for THz-enabled electric circuitry, IRS-assisted THz communication requires accurate signal modeling, which is different than the mm-Wave counterpart. 
	
	
	\subsubsection{Integrated sensing and communications (ISAC)}
	Recent research in ISAC envisions spectrum-sharing radar and communications in a hardware- and energy-efficient paradigm~\cite{irs_indoor_RIS_ISAC}. Here, again, IRS has been shown to allow range extension, NLoS sensing/communications, improved interference suppression, and enhanced security. DL-aided techniques have been identified for joint processing, multi-hop channel acquisition, and reduced post-training complexity for various ISAC tasks.
	
}

	\section{Summary and Future Outlook}
	\label{sec:Conc}
	We surveyed DL architectures of IRS-assisted wireless systems for key applications, including signal detection, channel estimation, and beamforming. We extensively discussed various learning architectures, such as SL, UL, FL, and RL, and their IRS-specific considerations. 
	
	{Compared to UL and RL, SL involves more complexity due to labeling process while it provides more accurate prediction/classification accuracy. FL is more efficient in terms of data/model transmission overhead and privacy as compared to CL. }
	
	While the label-free methods such as UL and RL have low complexity during training data generation, their performance suffers in comparison to the label-equipped SL. Note that the UL still requires an optimization stage for each data instance. The RL is promising because of its standalone operation and the consequent ability to adapt to environmental changes albeit at the cost of long training times.
	
	{The transmission overheads are significantly reduced in FL, which may be combined with other learning methods. For example, the combination of FL- and RL-based learning policies not only exhibits a communications-efficient model training but also provides environmental adaptation. Furthermore, hybrid FL and CL schemes can be new training strategies.}

\balance
\bibliographystyle{IEEEtran}
\bibliography{IEEEabrv,references_079_journal}
%
%


\end{document}